\documentclass[12pt,preprint]{emulateapj}

\slugcomment{To appear in Astrophysical Journal Letters}

\shorttitle{NGC~404 star-forming ring}
\shortauthors{Thilker et al.}

\begin{document}


\title{NGC~404, a rejuvenated lenticular galaxy\\on a merger-induced, blueward excursion into the green valley}


\author{David A. Thilker\altaffilmark{1}, Luciana Bianchi\altaffilmark{1}, David Schiminovich\altaffilmark{2}, Armando Gil de Paz\altaffilmark{3}, Mark Seibert\altaffilmark{4}, \\Barry F. Madore\altaffilmark{4}, Ted Wyder\altaffilmark{5}, R. Michael Rich\altaffilmark{6}, Sukyoung Yi\altaffilmark{7}, Tom Barlow\altaffilmark{5}, Tim Conrow\altaffilmark{5}, Karl Forster\altaffilmark{5}, Peter Friedman\altaffilmark{5}, Chris Martin\altaffilmark{5}, Patrick Morrissey\altaffilmark{5}, Susan Neff\altaffilmark{8}, and  Todd Small\altaffilmark{5}}
\email{dthilker@pha.jhu.edu}


\altaffiltext{1}{Center for Astrophysical Sciences, The Johns Hopkins University, 3400 N. Charles Street, Baltimore, MD 21218, USA}
\altaffiltext{2}{Department of Astronomy, Columbia University, 550 West 120th Street, New York, NY 10027, USA}
\altaffiltext{3}{Departamento de Astrof\'{i}sica y CC. de la Atm\'{o}sfera, Universidad Complutense de Madrid, Avda. de la Complutense, s/n, E-28040 Madrid, Spain}
\altaffiltext{4}{Observatories of the Carnegie Institution of Washington, 813 Santa Barbara Street, Pasadena, CA 91101, USA}
\altaffiltext{5}{California Institute of Technology, MC 405-47, 1200 East California Boulevard, Pasadena, CA 91125, USA}
\altaffiltext{6}{Department of Physics and Astronomy, UCLA, Los Angeles, CA 90095-1547, USA}
\altaffiltext{7}{Department of Astronomy, Yonsei University, Seoul 120-749, Korea}
\altaffiltext{8}{Laboratory for Astronomy and Solar Physics, NASA Goddard Space Flight Center, Greenbelt, MD 20771, USA}


\begin{abstract}
We have discovered recent star formation in the outermost portion
(1--4$\times$ $R_{25}$) of the nearby lenticular (S0) galaxy
NGC~404 using GALEX UV imaging.  FUV-bright sources are
strongly concentrated within the galaxy's HI ring (formed by a merger
event according to del Rio et al.), even though the average gas
density is dynamically subcritical.  Archival HST imaging reveals resolved upper main sequence stars and conclusively
demonstrates that the UV light originates from recent star formation
activity.  We present FUV, NUV radial surface brightness profiles and
integrated magnitudes for NGC~404.  Within the ring, the average
star formation rate surface density ($\Sigma_{\rm SFR}$) is
$\sim2.2\times10^{-5}$ M$_\odot$~yr$^{-1}$~kpc$^{-2}$.  Of the total FUV
flux, 70\% comes from the HI ring which is
forming stars at a rate of $2.5\times10^{-3}$ M$_\odot$~yr$^{-1}$. The
gas consumption timescale, assuming a constant SFR and no gas
recycling, is several times the age of the Universe.  In the context of the UV-optical galaxy CMD, the presence of the SF HI ring places NGC~404 in the green
valley separating the red and blue sequences. The rejuvenated
lenticular galaxy has experienced a merger-induced, disk-building
excursion away from the red sequence toward bluer colors, where it may
evolve quiescently or (if appropriately triggered) experience a burst
capable of placing it on the blue/star-forming sequence for up to
$\sim$1 Gyr.  The green valley galaxy population is heterogeneous,
with most systems transitioning from blue to red but others evolving
in the opposite sense due to acquisition of fresh gas through various channels.

%The evolutionary track of any
%particular galaxy is probably controlled by the quantity and means of
%gas supply, consumption, and removal.

\end{abstract}





\section{Introduction}

Strateva et al. (2001) used an early SDSS galaxy catalog to demonstrate
that the distribution of galaxies in color space is bimodal, and
correlated with galaxy morphology.  Galaxies populate
separate red and blue sequences in a ($u$ - $r$) versus $M_r$
color-magnitude diagram, CMD (Baldry et al. 2004).  Evolution of
galaxies from the blue sequence locus of actively star-forming systems to the red sequence occurs via transition
through an intermediate CMD zone, christened the green valley (Martin
et al. 2005).  Bell et al. (2004) and Faber et al. (2007) have both shown
that the red sequence has grown in mass over the period from $z\sim1$
to $z\sim0$.  Given that the color bimodality is driven principally by
star formation activity and UV bands are an excellent tracer of recent
star formation, it is not surprising that the galaxy color sequences
are especially well-separated in the UV-optical CMD (see Wyder et
al. 2007, hereafter W07).  GALEX proved instrumental in characterising the
incidence of green valley transition galaxies (W07; Martin et
al. 2007, hereafter M07) and understanding their evolution (M07, Schiminovich et al. 2007) from
star-forming systems (Salim et al. 2007) to ``red and dead'' galaxies.

The effect of gas removal, and the subsequent quenching of SF, has
been modeled (M07) in the context of galaxies evolving from the blue
sequence to the red sequence across the green valley.  However, the
influence of late addition of gas to red sequence galaxies has
received little attention. Contrary to classical expectations,
galaxies of the types predominately populating the red sequence
(elliptical and lenticular, S0) sometimes are associated with gaseous
reservoirs, especially if they are in a low density environment rather
than within a cluster (Morganti et al. 2008).  Oosterloo et al. (2007) showed the structure of such
gas is varied, with some regularly-rotating
disks and a complement of extended, offset, even tail-like
morphologies -- suggesting a diversity in origin.  Both galaxy mergers
and IGM accretion are viable mechanisms.  The presence of centralized
star formation in these early-type galaxies (ETGs) populating the red
sequence has been associated with low angular momentum sources (such
as retrograde mergers) because gas can be efficiently transported to
the remnant center and consumed (Serra et al. 2008), leaving behind
minimal HI.  On the other hand, prograde mergers of a red sequence
galaxy with a gas rich object may be the dominant mechanism of forming
massive, extended HI distributions around ETGs.

\object{NGC 404} is a prime target for studies addressing the structure and evolution of field S0 galaxies.  It is particularly interesting because it
is the nearest lenticular galaxy, lying just outside the Local Group
at an estimated distance of 3.1--3.3 Mpc (Karachentsev et al. 2002,
Tonry et al. 2001).  \object{NGC 404} is presently isolated in a zone
of radius $\sim$1.1 Mpc (Karachentsev \& Markarov 1999).  Although the galaxy
is not affected by a cluster environment (as is true for many
S0s), it may still have a complex history.  Karachentsev \& Markarov (1999) suggested \object{NGC 404} represents the end product
in the coalescence of a small galaxy group, having accreted with
all other (less massive) group members.  At least one merger may have
occurred in the recent past ($<$1 Gyr) and be fundamentally
responsible for the current UV morphology.  del Rio et al. (2004) argue that a large,
disk-like, neutral atomic hydrogen (HI) ring surrounding \object{NGC
404} is the remnant of a merger with a dwarf irregular galaxy which
took place some 900 Myr ago, according to their kinematic estimation.
Our GALEX observations revealed that this HI ring is now forming
stars.  However surprising, this rather late discovery is understandable post
facto since \object{NGC 404} is nearly hidden in the glare of red
supergiant Beta Andromedae (Mirach), 406$\arcsec$ to the
south east of the galaxy.  This bright foreground star
makes it difficult to image the faint surface brightness portions of
\object{NGC 404} at red wavelengths (including H$\alpha$), and the S0
galaxy became colloquially known as "Mirach's Ghost".

In this letter, we present evidence for star formation in \object{NGC
  404}'s outer disk-like HI ring.  We refrain from detailed interpretation
of UV properties of the inner disk and bulge. Section 2 describes our
GALEX observations.  Section 3 contains our data analysis.  We
conclude with discussion in Section 4.

\section{GALEX Observations}

%% In a manner similar to \objectname authors can provide links to dataset
%% hosted at participating data centers via the \dataset{} command.  The
%% second curly bracket argument is printed in the text while the first
%% parentheses argument serves as the valid data set identifier.  Large
%% lists of data set are best provided in a table (see Table 3 for an example).
%% Valid data set identifiers should be obtained from the data center that
%% is currently hosting the data.
%%
%% Note that AASTeX interprets everything between the curly braces in the 
%% macro as regular text, so any special characters, e.g. "#" or "_," must be 
%% preceded by a backslash. Otherwise, you will get a LaTeX error when you 
%% compile your manuscript.  Special characters do not 
%% need to be escaped in the optional, square-bracket argument.

We imaged NGC~404 with GALEX for 13463 seconds distributed
over 11 visits during Fall 2008 in both the far- and
near-ultraviolet (FUV, NUV) bands as part of the Nearby Galaxies Survey (NGS).  Our observations were coadded by
the GALEX pipeline.  The resulting deep FUV[NUV] image has
4.2[5.3]\arcsec resolution (Morrissey et al. 2007).  
%The rms noise in the FUV band is
%$2.7\times10^{-19}$ erg s$^{-1}$ cm$^{-2}$ \AA$^{-1}$ pixel$^{-1}$
%[29.0 ABmag/sq.arcsec].
% This corresponds to a limiting $\Sigma_{SFR}$ of ?? Ms/yr/kpc2 over what scale?.  

The FUV band of GALEX effectively suppresses glare from Mirach and
thus provides an unprecedented look at the star-forming stellar
population of \object{NGC 404}.  
%There are other advantages to
%observing in the UV (e.g. sensitivity to intermittent SF), and more
%specifically with GALEX (wide field-of-view) which have been detailed
%by Boissier et al. (2007) and Thilker et al. (2007).  
However, UV imaging can be susceptible to
extinction, effectively censoring the most embedded star formation
events (Calzetti et al. 2005, Thilker et al. 2007b).  In the outskirts
of \object{NGC 404} the expected
extinction is thought to be low overall, as we describe in Section 3.1.

Figure 1 (top) presents a side-by-side comparison of red POSS-II,
GALEX FUV, and HI (WHISP, van der Hulst et al. 2001) imaging for a
wide field centered on NGC~404, encompassing the galaxy's entire low
HI column density disk (detected until a radial distance of $\sim$
800$\arcsec$, 12.8 kpc).  The red light image gives no indication
regarding the presence of the HI disk. However, we find that the main
HI ring (between radii of 100--400$\arcsec$, 1.6--6.4 kpc) contains many FUV-bright
sources, far in excess of the number expected from the background
galaxy population.  These GALEX sources are frequently unresolved or
only marginally resolved, but often delineate larger aggregate
structures (sizes of $\sim$ 0.1 kpc to at least 2 kpc).  The large
scale UV structures are correlated with the most prominent areas of
HI, though sometimes with a positional offset between associated
features.
%This is particularly pronounced for the chain of UV-bright
%sources delineating the northern limit of the SF HI ring, running
%through $\delta$=+35:48:30.  
The UV/HI morphology is suggestive of spiral waves triggering the star
formation.  We also note that at least in the UV, the most significant
HI peaks do not show as prominent SF sites.  Some of the high
$N(HI)$ regions may be on the verge of forming the next generation of
stellar clusters, but haven't yet.  We have plans to check
for molecular gas concentrations matching the $N(HI)$ peaks.

We display a color-composite GALEX image of NGC~404 in the bottom
panel of Fig. 1.  The inner disk and bulge of NGC~404 appear
prominently in the NUV-band, with average $FUV-NUV$ = 1.8 ABmag within
$R_{25}$ (102$\arcsec$), as is typical for an S0 galaxy (Donas et
al. 2007).  The inner disk shows minimal substructure at 5\arcsec (80
pc) GALEX resolution, though other studies have described dust clouds
(Tikhonov et al. 2003), molecular gas (Sage 1990, Wiklind \& Henkel
1990) and an ionized disk (Plana et al. 1998) within 200 pc of the
LINER nucleus.  The single most luminous FUV source in the galaxy
occupies the nuclear area and is known to contain one dominant and
several faint clusters of massive stars (Maoz et al. 1995, 1998).  The
most remarkable aspect of the color image in Fig. 1 is the
distribution of blue (FUV-bright) sources occupying the main SF HI
ring. Typical FUV luminosity of individual clumps in this ring is
$1.5\times10^{38}$ erg s$^{-1}$ ($m_{FUV}$ = 23.0 ABmag, or
$M_{FUV}$ = -5.6 at 3.3 Mpc). All UV-related measurables in this
paper have been corrected only for Milky Way extinction, with $E(B-V)$
= 0.059 mag (Schlegel et al. 1998).  Comparing intrinsic UV
color to instantaneous burst Starburst99 models (Leitherer et al. 1999) we estimate most sources have age 30 Myr -- 1 Gyr and
corresponding stellar mass 10$^3$--10$^4$ M$_{\odot}$.
% hence $A(FUV)$ = 0.48 and $A(NUV)$ = 0.54. 
There is a sharp decline to the surface density of GALEX sources
having blue $FUV-NUV$ beyond a radius of 400\arcsec.  The $N(HI)$
radial profile of del Rio et al. (2004) indicates the average column
density in this zone beyond the main ring is $\sim4\times10^{19}$
cm$^{-2}$.  Two areas of exceptionally blue, yet diffuse FUV are seen
-- NE of the SF ring and overlapping with the ring to the
ESE. Their position within a larger filamentary network covering the
GALEX field of view suggests they are attributable to Galactic
cirrus.

\section{Analysis and Results}

\subsection{Global structure}

The top panel of Fig. 2 shows background-subtracted,
foreground-extinction corrected UV surface brightness profiles and
growth curves measured for NGC~404 after masking of foreground stars
and distant, unrelated galaxies.  The bottom panel of Fig. 2 presents
$FUV-NUV$ as a function of galactocentric distance.  Table 1 lists
integrated magnitudes and several
quantities derived after folding in corollary data.

We neglect any average internal extinction, supported by
inspection of Spitzer 24$\mu$m imaging of the SF ring obtained as part
of the LVL Survey (Lee et al. 2009, Dale et al. 2009).  Very few of the FUV-bright
sources in NGC 404's ring have candidate 24$\mu$m counterparts.
Indeed, association in these cases may result from confusion of
dusty background galaxies with UV ring clumps. Our extinction
assumption is also supported by Prescott et al. (2007) who show that
the fraction of highly obscured SF regions in ordinary galaxy disks is
very small ($4\%$).  Because we focus on a comparatively low
density, outer-disk environment, this is likely even more true
in the NGC~404 ring (with at most a few significantly obscured SF
complexes).

Fig. 2 shows that: (1) the main HI ring, at galactocentric distance
$R$=100-400$\arcsec$, has an average $\mu_{FUV} \sim 29$
ABmag/sq.arcsec; (2) $FUV-NUV$ appears to slowly decline with
increasing radius in the ring, from about 2 to 0.2, probably
reflecting contamination from an underlying exponential disk at least
for smaller radii; (3) UV color for the inner disk/bulge is nearly
constant at $FUV-NUV$=2.2 except for the nuclear area where FUV
surface brightness increases (perhaps due to the LINER, or stronger
UV-upturn in central regions of the bulge, e.g. Ohl et al. 1998),
bringing $FUV-NUV$ down to 1.1.  From the curve of growth, we
determine that 70\% of the total FUV
luminosity ($L(FUV)$ = $1.0\times10^{41}$ erg~s$^{-1}$,
corresponding to $FUV$=14.9) in NGC~404 comes from the SF HI ring
defined in (1).  The HI ring has a SFR of 0.0025 M$_{\odot}$
yr$^{-1}$, adopting the SFR calibration of Salim et al. (2007, their
Eq. 8) for a Chabrier IMF over 0.1-100 M$_{\odot}$.

%FUV-NUV of 0.2 corresponds to SSP age of approx. 300 Myr -- whereas
%values approaching 2.0 are about 1 Gyr, but contamination and
%breakdown of SSP assumption.  UV-vis stellar photom to give answer,
%and timescale for merger.  At the very least FUV-NUV are consistent
%with del Rio hypothesis.  0.2 is too red for CSP models of FUV-NUV,
%even at solar met.  hmm? what about ongoing SF given the gas mass?

\subsection{HST detection of a resolved young, stellar population}

Very deep F606W (39000s) and F814W (75400s) WFPC2 imaging of NGC~404
was obtained as part of ANGST (Dalcanton et al. 2009) and
serendipitously samples a subsection of the SF ring
($R \sim$ 160$\arcsec$, 2.6 kpc).  We obtained the ANGST public-release
photometric catalog of resolved stars from the Hubble Legacy Archive
(HLA).  Coadded images in F606W and F814W bands were retrieved from
the ST-ECF HST Archive.  At the color of the main
sequence, ANGST observations of NGC~404 are 100\% complete for sources
brighter than F814W = 25.5 mag and 50\% complete down to $\sim$26.7
mag.  This depth is sufficient to recover all massive main sequence
stars.

%Figure 3 displays a color representation of the F606W, F814W WFPC2
%imaging from ANGST with FUV source boundaries overlaid (Thilker et
%al. 2009, in prep.).  The nuclear area of NGC~404 lies to the SE,
%outside the field of view.  

The NGC~404 ANGST dataset will be described in a dedicated paper
(Thilker et al. in prep.) but we note the following after initial
image inspection: (1) there is minimal surface density enhancement
within UV clumps, consistent with a single star or few stars emitting
significant UV (cf. Gil de Paz et al. 2007); (2) occasional evidence
for diffuse, nebular (H$\alpha$) emission in the F606W band is seen;
(3) the overall distribution of resolved stars declines in surface
density as a function of galactocentric distance, indicating the
continued presence of an older, evolved disk (noted by Tikhonov et
al. (2003) for a different WFPC2 field) at large $R$ underlying the
ring; and (4) a small minority of the UV clumps may be contaminating
sources such as background galaxies.

The HST data confirm that the UV light detected by GALEX originates from
a young stellar population.  
%The righthand portion of Fig. 3 presents
%a CMD of resolved stars in the WFPC2 field, indicating those within UV
%clumps using blue circles and all other stars as dots.  Overlaid on
%the CMD are reddened theoretical isochrones from Marigo et al. (2008).
%They were computed for an estimated metallicity of NGC~404, [Fe/H] =
%-0.6 (see below), and reddened in accord with the foreground
%extinction. 
% of A$_V$ = 0.194 mag (Schlegel et al. 1998).  
%Selected
%masses on the ZAMS are also marked. 
CMD analysis of the ANGST catalogs in comparison to theoretical isochrones (Marigo et al. 2008) shows that in the UV clumps we detect main
sequence stars ranging in mass from $\sim$5--40 M$_{\odot}$, post
turn-off stars (including blue loop), and a background of RGB/AGB
stars.  Post turn-off stars in the 100-300 Myr age range are also
found outside of the UV clumps, suggesting stars from dispersed
clusters may be present in the field.

%  A detailed CMD analysis of the
%SFH in this WFPC2 dataset could directly probe the timescale for the
%merger and its effect, but is beyond the scope of this letter.  

%We
%note that based on the color of RGB stars, the ANGST data are more
%compatible with a range in metallicity (roughly [Fe/H] from -1 to
%-0.3, with few stars at the low end) higher than that determined by
%Tikhonov et al. (2003, [Fe/H]=-1.1).

\subsection{Star formation law \& threshold considerations}

del Rio et al. (2004) showed that gas surface density in the NGC~404
ring is subcritical in terms of the azimuthally-averaged $Q$ parameter.  Our data indicate that SF occurs in this environment, having average $\Sigma_{HI}$ = 1.2
M$_{\odot}$ pc$^{-2}$ between $R$=100-400$\arcsec$.  This is
not surprising given other low density SF detected by GALEX
(XUV-disks, Thilker et al. 2007; ordinary spirals, Boissier et
al. 2007; low-surface-brightness galaxies, Wyder et al. 2009; ETGs, Donovan et al. 2009).  Local
conditions must still be conducive to massive star formation. Indeed, clumps in the HI ring do reach $\Sigma_{HI}$ $\sim$ 4 M$_{\odot}$ pc$^{-2}$.  

%%1.45-1.65d20 cm$^2$ between 150-300\arcsec radius
%%1.2 Msun pc-2, 5x too low for eff. SF, 10x subcritical dyn.

However, the average SFR surface density in NGC~404's ring
($\Sigma_{SFR}$ $\sim$ $2.2\times10^{-5}$
M$_\odot$~yr$^{-1}$~kpc$^{-2}$) is remarkably low and provides further
constraint on the SF law at low density.  Even assuming negligible
molecular content, the NGC~404 ring falls 1 dex below the traditional
Kennicutt (1998) (log $\Sigma_{SFR}$, log $\Sigma_{gas}$) correlation.
The implied gas consumption timescale is $\sim 4\times t_H$ (Hubble
time, 1/$H_0$), without gas recyclying. Equivalently, we infer SF
efficiency of $1.7\times10^{-11}$ yr$^{-1}$ (0.16\% over 100 Myr, or
$>20\times$ less than typical in galactic disks).
%Bigiel08 Fig 8 shows 3.4% on FUV+24 vs HI+H2 panel for H2 dominated, Leroy give 5.25d-11 for H2 alone)
The NGC~404 data is roughly consistent with the theoretical SF law of
Krumholz et al. (2009), including a downturn in SF efficiency at low
density due to a suppressed molecular fraction.  The
remaining offset (although small) is consistent with not accounting
for a small contribution of molecular gas in the cloud cores of the
ring -- it must be comparatively rare, but still present.  This also
true for the Wyder et al. (2009) sample of LSB galaxies.

%0.079 log gas using 1.2 value below
%-4.66 for log SigSFR

\subsection{Backwards evolution into the green valley}

In Figure 3 we plot the logarithm of volume density of galaxies in the
UV-optical galaxy CMD. W07 fits to the red and
blue (M$_r$, $NUV-r$) sequences are marked.  The green valley is the
transitional region between these sequences.  Conventionally most
green valley objects are thought to be evolving from the blue to the
red sequence, after cessation of star formation activity.  The GALEX
photometry presented above is used to mark the position of NGC~404 in
the galaxy CMD, both with (green) and without (red) the contribution
of the HI ring.  Optical magnitudes in both cases were determined by
extrapolating the $r$-band, 84.6$\arcsec$ diameter aperture
measurement of Sandage \& Visvanathan (1978) with the $I$-band surface
brightness profiles of Tiknohov et al. (2003).  

Assuming NGC~404's HI ring originates from a recent merger
(del Rio et al. 2004), then the galaxy has evolved from a red sequence
object and moved backwards into the green valley. We do not yet have a
detailed recent SFH to know if is already returning to the red
sequence or whether it will continue to evolve bluewards.  At the
present SFR, it could indefinitely continue quiescently building
(SFR/M$_{*}$ = -11.4) a low surface brightness outer disk.
Alternatively, if an epoch of enhanced SF were triggered perhaps by
another encounter, consumption of the gas reservoir could position the
galaxy on the blue/S-F sequence (SFR/M$_{*}$=-9.4) for $\sim$1 Gyr.
If the available HI in the ring were eventually all converted into
stars, it could contribute an additional $\sim$20\% to the stellar
mass of NGC404 (currently $6.9\times10^8$ M$_{\odot}$ derived from SED fitting
of UV-NIR integrated photometry), while  substantially decreasing the bulge-to-disk mass ratio.

%$1.09\times10^9$ M$_{\odot}$, derived from K$_s$ = 7.53 measurement of
%Jarrett et al. 2003 and the calibration of Leroy et al. 2008),
%boosting the importance of the disk relative to the bulge.

%(now B/D mass ratio = ? ... after?).  

%%total mass from Keplerian fit to rot curve - 3d10, how much baryonic though? del rio poster argues no dm beyond ring

%%2mass lga total k$_s$ (r<116) = 7.532+-0.024 --> convert to mass [binney \& merrifield 98 -- kmag of sun =3.28

%%%OLD DISTANCE....IDL> print,10.d0^(0.4*((3.28+(5.*alog10(3.13d6)-5.))-7.53))*0.5= 9.77e+08 = 1.d9 msun
%%%ADOPTED DISTANCE: IDL> print,10.d0^(0.4*((3.28+(5.*alog10(3.3d6)-5.))-7.53))*0.5
%%%   1.0864204e+09 msun
%%%IDL> print,1.5d8/1.08d9
%%%      0.13888889  ~15% increment to stellar mass if all HI converted to stars

\section{Discussion}

Star formation and associated gas are no longer viewed as uncommon in
lenticular and elliptical galaxies, owing to ever more sensitive UV
and HI imaging surveys.  Residual star formation is thought to be
present in 10--30\% of the early-type galaxies examined with GALEX
(Yi et al. 2005, Donas et al. 2007, Schawinski et al. 2007, Kaviraj et al. 2007), even after excluding classical UV upturn candidate galaxies from the sample.  An external source of gas
(accretion or mergers) very likely supplements the SF fuel available
via recycling from stellar mass loss.  HI imaging for ETG
samples supports this conclusion, frequently revealing extended
gaseous distributions (Morganti et al. 2008). What remains to be
determined is the long term effect of externally-fueled SF on the
morphology of red-and-dead galaxies and on the appearance of the UV-optical galaxy
CMD.  Sometimes the gas is quickly consumed near the galaxy center
(Serra et al. 2008), having little net effect on morphology and a very
short-lived movement within the galaxy CMD.  Activity of this sort is
the ``frosting'' variety described by Trager et al. (2000), though
Schawinski et al. (2009) also describe ETGs with rather high centralized
SFRs ($>50$ M$_{\odot}$ yr$^{-1}$).  We have shown that rejuvenation
of large-scale disk formation is another possible outcome, as
discovered in NGC~404 at a very low level.  NGC~404 is classified as a Type 1 XUV-disk using the criteria of Thilker et al. (2007). Similar cases have been reported by Cortese \& Hughes (2009), Donovan et al. (2009) and Rich \& Salim (2009). Rejuvenation events may effectively
transplant red sequence galaxies to the green valley and blue sequence.
%quasi-permanently
  We note that E/S0 galaxies have been detected already on the blue
sequence (Kannappan et al. 2009, Schawinski et al. 2009).  NGC~404 represents a possible example of
this transition underway.  

%%not the only galaxy like this seen by GALEX... 

We conclude that the galaxy CMD transition zone known as the green
valley represents a heterogeneous population of objects, with many
evolving from blue to red but others going in the opposite direction.
Stochastic excursions into the green valley from the red-sequence,
driven by acquisition of fresh gas for star formation, are observed.
Traffic through the green valley is not one-way.  NGC~404 confirms
that the merger of ETGs and gas-rich dwarfs are one mechanism
establishing this diversity.  This implies UV imaging is fundamentally
required to place individual ETGs into an accurate evolutionary
context, assessing whether they are red and dead (at least for the
moment), experiencing residual star formation in the galaxy center, or
entering a rejuvenated phase of disk building fueled by a long-lasting
gas reservior.  In fact, UV imaging is an effective means of detecting
such potentially transformational reservoirs.


%% Included in this acknowledgments section are examples of the
%% AASTeX hypertext markup commands. Use \url without the optional [HREF]
%% argument when you want to print the url directly in the text. Otherwise,
%% use either \url or \anchor, with the HREF as the first argument and the
%% text to be printed in the second.

\acknowledgments
GALEX (Galaxy Evolution Explorer) is a NASA Small Explorer, launched
in April 2003. We gratefully acknowledge NASA's support for
construction, operation, and science analysis for the GALEX mission,
developed in cooperation with the Centre National d'Etudes Spatiales
of France and the Korean Ministry of Science and Technology.  This research has made use of the NASA/IPAC Extragalactic Database
(NED).  We acknowledge the usage of the HyperLeda database
(http://leda.univ-lyon1.fr).  The Digitized Sky Surveys were produced
at the Space Telescope Science Institute under U.S. Government grant
NAG W-2166.   Some images presented in this paper
were obtained from the Multimission Archive at the Space Telescope
Science Institute (MAST).


%% in parentheses, after the keyword, but they will not be verified.

{\it Facilities:} \facility{GALEX}, \facility{HST (WFPC2)}, \facility{NRAO (VLA)}, \facility{WSRT}.




\clearpage

%% the file, uncomment the \plotone, \plottwo, and
%% If you need a layout that cannot be achieved with \plotone or
%% \plottwo, you can invoke the graphicx package directly with the
%% \includegraphics command or use \plotfiddle. more information,
%% please see the tutorial on "Using Electronic Art with AASTeX" in the
%% documentation section at the AASTeX Web site,
%% http://www.journals.uchicago.edu/AAS/AASTeX.
%%
%% The examples below also include sample markup for submission of
%% supplemental electronic materials. As always, be sure to check
%% the instructions to authors for the journal you are submitting to
%% for specific submissions guidelines as they vary from

%% This example uses \plotone to include an EPS file scaled to
%% 80% of its natural size with \epsscale. Its caption
%% has been written to indicate that additional figure parts will be
%% available in the electronic journal.

\begin{figure}
%\epsscale{.80}
\plotone{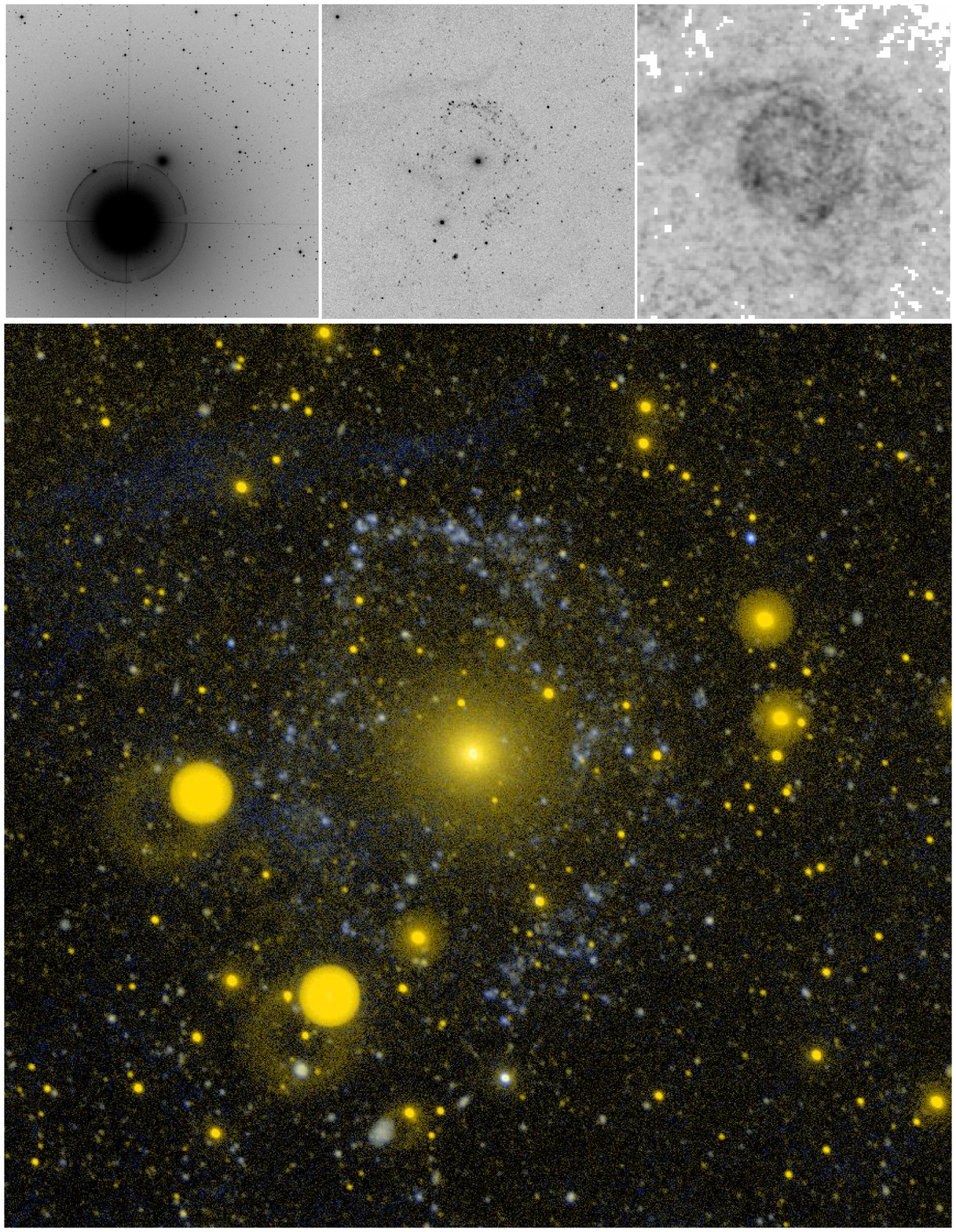}
\caption{DSS2-red ({\em top-left}), GALEX FUV ({\em top-center}), HI ({\em top-right}) imaging of a wide-field centered on NGC~404, plus a detailed FUV, NUV color-composite image of NGC~404 ({\em bottom}).   The FUV ($\lambda_{eff}$ = 1538.6 \AA) imaging reveals star formation in the ring.    The wide-field panels (at top) span 0.5\degr$\times$0.5\degr (27.3 kpc) with N up and E left.  The GALEX color-composite was created by assigning FUV to the blue channel, NUV ($\lambda_{eff}$ = 2315.7 \AA) to the red channel, and an energy-weighted average of FUV and NUV to green channel. FUV-bright sources appear blue/white, including the majority of the detections in the SF ring. The inner disk and bulge appears yellow because it is NUV-bright, dominated by older stars than the SF ring.  Dichroic reflections appear as yellow doughnuts of varied size near bright foreground stars (including Mirach, SE of the galaxy). The color-composite field is 22.85\arcmin$\times$21.80\arcmin (20.8$\times$19.8 kpc).}
\end{figure}

\begin{figure*}
%\plotone{/home/dthilker/ngc404allprof.ps}
\plotone{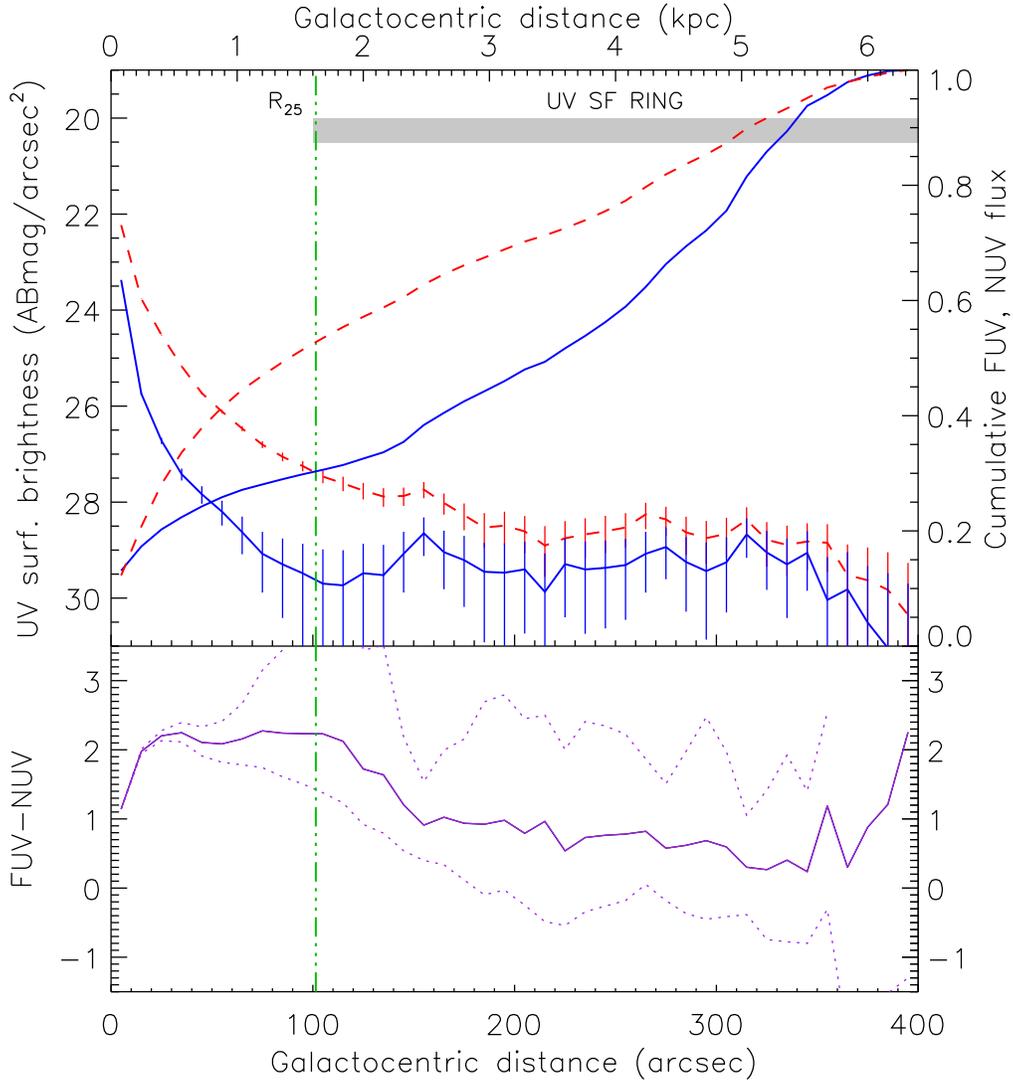}
\caption{GALEX radial profile analysis for NGC~404.  ({\em top}) Average surface brightness and corresponding curve of
growth as a function of galactocentric distance for the FUV (blue
solid) and NUV (red dashed) GALEX bands.  The $\pm$1$\sigma$ confidence intervals for the surface brightness profiles, including the uncertainty from sky subtraction, are plotted with thin color-coded lines.  ({\em bottom}) Azimuthally-averaged FUV-NUV color with error limits (dotted lines), which are dominated by sky level uncertainty.}
\end{figure*}

%\begin{figure*}
%\epsscale{.80}
%\plottwo{/data3/marmarole/dthilker/papers/ngc404letter/n404f814wreg_test.ps}{/data3/marmarole/dthilker/papers/ngc404letter/n404testcmd.ps}
%\caption{(left) HST WFPC2 imaging for a subsection of the NGC~404 ring, with FUV clump boundaries overlaid. (right) CMD for resolved stars, with Marigo et al. (2008) isochrones marked (purple to red, 10, 50, 100, 300, 500 Myr, 1 , 5, 10 Gyr).  Stars within FUV clumps are marked with circles.  Positions for ZAMS stars of 5, 10, 20, 50 M$_{\odot}$ are indicated with triangles.  }
%\end{figure*}  

\begin{figure*}
\epsscale{.80}
%\plotone{/data3/marmarole/dthilker/papers/ngc404letter/galaxycmd.ps}
\plotone{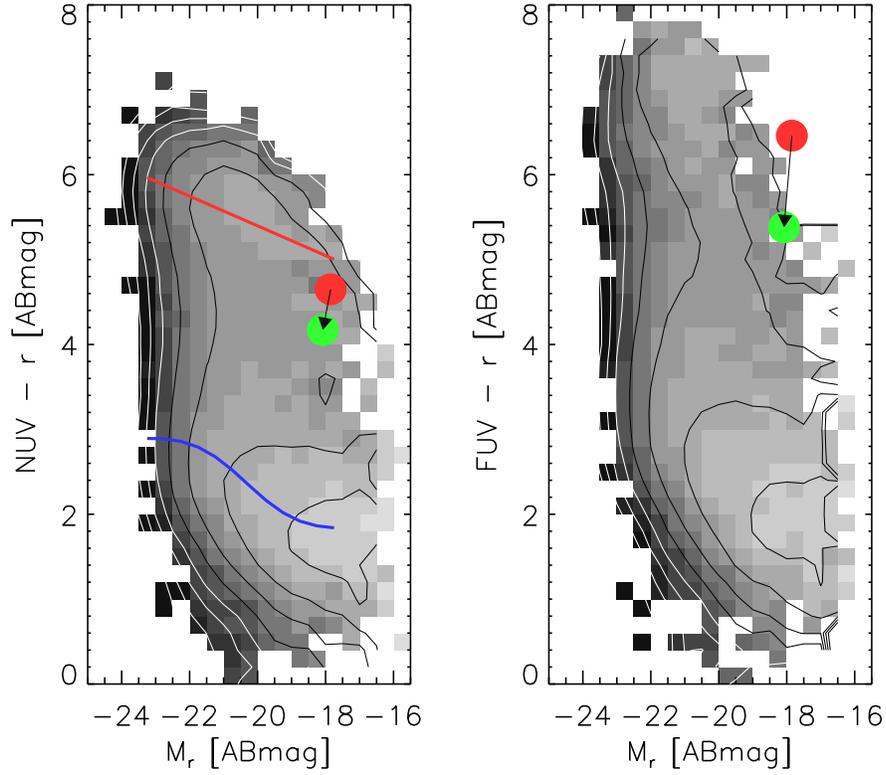}
\caption{UV-optical galaxy CMDs [{\em left}, (M$_r$,$NUV-r$); {\em right}, (M$_r$,$FUV-r$)]
adapted from W07, upon which we have marked the position of NGC~404
both with and without the contribution of flux from SF HI ring,
using green and red dots respectively. In the (M$_r$,$NUV-r$) CMD we overplot the W07 fits to the red and blue galaxy sequences.  NGC~404 falls into the green valley when the ring is included. Galaxy volume density is greyscaled over a range of 6 dex, with contours drawn every 0.5 dex.}
\end{figure*}

%% If you are not including electonic art with your submission, you may
%% mark up your captions using the \figcaption command. See the
%% No more than seven \figcaption commands are allowed per page,
%% so if you have more than seven captions, insert a \clearpage

\clearpage

\begin{deluxetable}{lccc}
%%\tabletypesize{\scriptsize}
\tablecaption{Properties of NGC~404}
\tablewidth{0pt}
\tablehead{
\colhead{Quantity} & \colhead{Value} & \colhead{Unit} & \colhead{Ref}}
\startdata
Hubble Type& SA(s)0- &&(1)\\
Morph. Code, $t$ & -2.8$\pm$0.6 &&(2)\\
Spectral (AGN) Type & LINER & &(1)\\
RA (J2000)& 01h 09m 27.010s&&(1)\\
DEC (J2000)&+35d 43m 04.20s&&(1)\\
Heliocentric Velocity & -48$\pm$9 & km s$^{-1}$&(1)\\
Distance\tablenotemark{a}  & 3.3 & Mpc&(3)\\
$D_{25}$& 203$\pm$14&arcsec&(2)\\
$D_{HI}$& $\sim1600$&arcsec&(3)\\
Inclination, $i$& 0--11&deg&(2,3)\\
$M_{tot}$&$3\times10^{10}$&M$_{\odot}$&(3)\\
$M_{stellar}$&$6.9\times10^{8}$&M$_{\odot}$&(4)\\
$M_{HI}$&$1.5\times10^{8}$&M$_{\odot}$&(3)\\
$M_{tot}/L_B$&85&M$_{\odot}$/L$_{\odot}$&(3)\\
$M_{HI}/L_B$&0.42&M$_{\odot}$/L$_{\odot}$&(3)\\
DEF(HI), HI deficiency\tablenotemark{b}&-0.54&&(3)\\
$M_{HI}/M_{stellar}$&0.22&&(3,4)\\
Galactic $E(B-V)$&0.059&mag&(1)\\
%%%note that L_B is in units of Lsun,bol not Lsun,B
%%%print,10.d0^(0.4*(4.72-(10.94-27.59))=3.53d8
%%%where Lsun,bol=4.72, H.L. btc for N404 = 10.94, m-M=27.59 for 3.3Mpc  
%%%Mtot/LB = 3.d10/3.53d8 = 85
%%%MHI/LB = 1.5d8/3.53d8 = 0.42
\\
\hline
\\
$FUV\tablenotemark{c}$ & 14.89 & ABmag & (4)\\
$NUV$ & 13.68 & ABmag & (4)\\
$r$ & 9.51 & ABmag & (5,6)\\
$FUV(R<100\arcsec)$ & 16.20 & ABmag & (4)\\
$NUV(R<100\arcsec)$ & 14.39 & ABmag & (4)\\
$r(R<100\arcsec)$ & 9.74 & ABmag & (5,6)\\
$L(FUV)$&$1.0\times10^{41}$&erg s$^{-1}$& (4)\\
$SFR_{tot}$ & $3.6\times10^{-3}$ & M$_{\odot}$ yr$^{-1}$&(4)\\
$SFR_{ring}$ & $2.5\times10^{-3}$ & M$_{\odot}$ yr$^{-1}$&(4)\\
$\Sigma_{SFR,ring}$&$2.2\times10^{-5}$&M$_{\odot}$ yr$^{-1}$&(4)\\
Gas consumption time, $\tau$ &59&Gyr&(4)\\
$SFE_{HI}$&$1.7\times10^{-11}$& yr$^{-1}$&(4)\\

\enddata
%% Text for table notes should follow after the \enddata but before
%% the \end{deluxetable}. Make sure there is at least one \tablenotemark
%% in the table for each \tablenotetext.
\tablecomments{(1) NED, (2) HyperLEDA, (3) del Rio et al. (2004), (4) this paper, (5) Sandage \& Visvanathan (1978), (6) Tikhonov et al. (2003)}
\tablenotetext{a}{We adopted 3.3 Mpc for consistency with del Rio et al. (2004).}
\tablenotetext{b}{Based on the Haynes \& Giovanelli (1984) definition.}
\tablenotetext{c}{All quantities below have been corrected for foreground extinction using the Cardelli et al. (1989) extinction law.} 
\end{deluxetable}

%% The following command ends your manuscript. LaTeX will ignore any text

\end{document}